\newif\ifsubmode
\submodefalse

\newif\ifprintfig
\printfigtrue

\ifsubmode
  \documentstyle[12pt,aasms4,epsf]{article}
  \received{}
  \accepted{}
  \journalid{}{}
  \articleid{}{}
\else
  \documentstyle[11pt,aaspp4,epsf]{article}
  \slugcomment{{\it Astronomical Journal, in press (January 2003)}}
\fi

\lefthead{Gerssen et al.}
\righthead{Addendum: ``HST Study of M15 --- II''}

\newcommand{\etal}{{et al.~}}

\newcommand{\Msun}{\>{\rm M_{\odot}}}

\begin{document}

\title{Addendum: ``Hubble Space Telescope Evidence for an Intermediate-Mass 
Black Hole in the Globular Cluster M15---\\ II.~Kinematical Analysis
and Dynamical Modeling\altaffilmark{1}''}

\author{Joris Gerssen, Roeland P.~van der Marel}
\affil{Space Telescope Science Institute, 3700 San Martin Drive,
       Baltimore, MD 21218}

\author{Karl Gebhardt}
\affil{Department of Astronomy, Mail Code C4100, University of Texas at 
Austin, Austin, TX 78712}

\author{Puragra Guhathakurta\altaffilmark{2,3}}
\affil{Herzberg Institute of Astrophysics, National Research Council of Canada,
5071 West Saanich Road, Victoria, BC V9E 2E7, Canada}

\author{Ruth C.~Peterson\altaffilmark{4}}
\affil{UCO/Lick Observatory, Department of Astronomy and Astrophysics,
       University of California at Santa Cruz, 1156 High Street,
       Santa Cruz, CA 95064}

\author{Carlton Pryor}
\affil{Department of Physics and Astronomy, Rutgers University, 136
       Frelinghuysen Road, Piscataway, NJ 08854-8019}


\altaffiltext{1}{Based on observations made with the NASA/ESA Hubble Space
Telescope, obtained at the Space Telescope Science Institute, which is
operated by the Association of Universities for Research in Astronomy,
Inc., under NASA contract NAS 5-26555. These observations are
associated with proposal \#8262.}

\altaffiltext{2}{Herzberg Fellow.}

\altaffiltext{3}{Permanent address: UCO/Lick Observatory, University of 
California, Santa Cruz, 1156 High Street Santa Cruz, California 95064.}

\altaffiltext{4}{Also at: Astrophysical Advances, Palo Alto, CA 94301.}


\ifsubmode\else
\clearpage\fi


\ifsubmode\else
\baselineskip=14pt
\fi


In our paper we reported the existence of a dark and compact mass
component near the center of M15, based on an analysis of new data
from the Hubble Space Telescope. Possible explanations for this mass
component include: (1) a single intermediate-mass black hole (BH); or
(2) a collection of dark remnants (e.g., neutron stars) that have sunk
to the cluster center due to mass segregation. We assessed the
plausibility of the latter possibility by comparing the kinematical
data for M15 to the predictions of the most sophisticated and most
recently published Fokker-Planck models for M15 (Dull
\etal 1997, hereafter D97). We showed that the mass-to-light ratio ($M/L$) 
profile in Figure~12 of D97 implies too few dark remnants near the
cluster center to explain the observed kinematics of M15. This
supported the view that M15 harbors an intermediate-mass BH. We
address here how this conclusion is affected by the recent discovery
of an error in Figure~12 of D97. We show that the presence of an
intermediate-mass BH continues to be a viable interpretation of the
data, but that its presence ceases to be uniquely implied.

After the completion of our paper, the authors of the D97 paper
discovered an unfortunate error in their Figures~9 and~12. The
labeling along the abscissa of these figures is incorrect due to a
coding error in SM plotting routines (H.~Cohn and B.~Murphy, private
communication, 2002). The units along the top axis should have read
`arcmin' instead of `pc', and the labeling in arcmin along the bottom
axis is incorrect. The net result is that the radial scale of these
figures in the D97 paper is too compressed by a factor $2.82$. The
true total mass of the centrally concentrated population of dark
remnants in the Fokker-Planck models is therefore considerably larger
than what was implied by the $M/L$ profile shown in Figure~12 of the
D97 paper.

Figure~1 shown here is similar to Figure~12 of our paper, but it now
shows the data-model comparison with the corrected D97 $M/L$ profile.
Models without a BH are now found to be statistically acceptable
(within $1 \sigma$), although inclusion of an intermediate-mass BH,
with $M_{\rm BH} = 1.7^{+2.7}_{-1.7} \times 10^3 \Msun$, still
provides a marginally better fit to the data (although not a
statistically significant one). Hence, the Fokker-Planck models for
M15 discussed by D97 do in fact have enough dark remnants near the
cluster center to explain the observed kinematics. However, the D97
models still have a number of important shortcomings, as discussed in
\S5.4 of our paper. Most importantly, D97 assumed that all neutron
stars that form in the cluster are retained. By contrast, the observed
distribution of pulsar kick velocities indicates that the retention
factor should only be a few percent; most authors agree that it should
be no more than 10\% (see references in \S5.4 of our paper). In this
sense, the D97 models provide an upper limit on the number and mass of
dark remnants in M15. The same is true for $N$-body models constructed
recently by Baumgardt \etal (2002), the results of which are
qualitatively similar to those of D97. More realistic evolutionary
models that include neutron star escape will require a more massive
and more statistically significant BH to fit the data than that
suggested by Figure~1. Alternatively, one can assume that there are
more stars in the high-mass end of the initial mass function, or that
the transition between stars that evolve to white dwarfs as compared
to neutron stars occurs at a higher initial mass (H.~Cohn and
B.~Murphy, private communication, 2002). Independent evidence does not
exist to support these assumptions, although they cannot be ruled out.

The evidence for a central BH in M15 is less convincing than it was on
the basis of our earlier analyses. However, because of the somewhat
unrealistic assumptions about neutron star retention in the models of
D97 and Baumgardt \etal (2002), and because of the independent
evidence for a BH in another cluster (G1; Gebhardt, Rich \& Ho 2002),
the presence of a BH in M15 continues to be a viable interpretation of
the data. The best fit BH mass with the corrected D97 $M/L$ profile is
$M_{\rm BH} = 1.7^{+2.7}_{-1.7} \times 10^3 \Msun$ (see Figure~1);
with a constant $M/L$ it is $M_{\rm BH} = 3.2^{+2.2}_{-2.2} \times
10^3 \Msun$ (see \S5.4 of our paper). A model that includes both
neutron star escape and mass segregation would probably yield a value
between these numbers. So if M15 has a BH, its mass is consistent with
the correlation between velocity dispersion and BH mass that has been
inferred for galaxies (see Figure~14 of our paper). This continues to
suggest the possible existence of an important new link between the
structure, evolution and formation of globular clusters, galaxies, and
their central BHs (see \S6 of our paper, and also Gebhardt \etal
2002). However, with the presently available models and data it is
neither uniquely implied nor ruled out that M15 has an
intermediate-mass BH.


\acknowledgments

We thank Haldan Cohn, Brian Murphy, Phyllis Lugger, Piet Hut and Steve
McMillan for stimulating discussions.


\ifsubmode\else
\baselineskip=10pt
\fi



\clearpage


\ifsubmode\else
\baselineskip=14pt
\fi


\newcommand{\figcapmaxlike}{Data-model comparison for spherical dynamical
models with an isotropic velocity distribution, an $M/L$ profile
inferred from Fokker-Planck models, and a central black hole of mass
$M_{\rm BH}$. The $M/L$ profile is a corrected version of the one
shown by D97. {\bf (a; left panel)} The likelihood quantity $\lambda$
defined in equation~(4) of our paper as function of $M_{\rm BH}$. The
minimum in $\lambda$ identifies the best fit black hole
mass. Horizontal dashed lines indicate the $1$ and $2\sigma$
confidence regions. {\bf (b; right panel)} The RMS projected
line-of-sight velocity $\sigma_{\rm RMS}$ as a function of projected
radius $R$. The heavy jagged curve surrounded by heavy dashed curves
is the observed profile, as in Figure~9d of our paper. The smooth thin
curves are the predictions for models with $M_{\rm BH}$ ranging from 0
to $10 \times 10^3 \Msun$ in steps of $10^3 \Msun$.\label{f:maxlike}}


\ifsubmode
\figcaption{\figcapmaxlike}
\clearpage
\else\printfigtrue\fi

\ifprintfig


\setcounter{figure}{0}

\clearpage
\begin{figure}
\epsfxsize=0.9\hsize
\centerline{\epsfbox{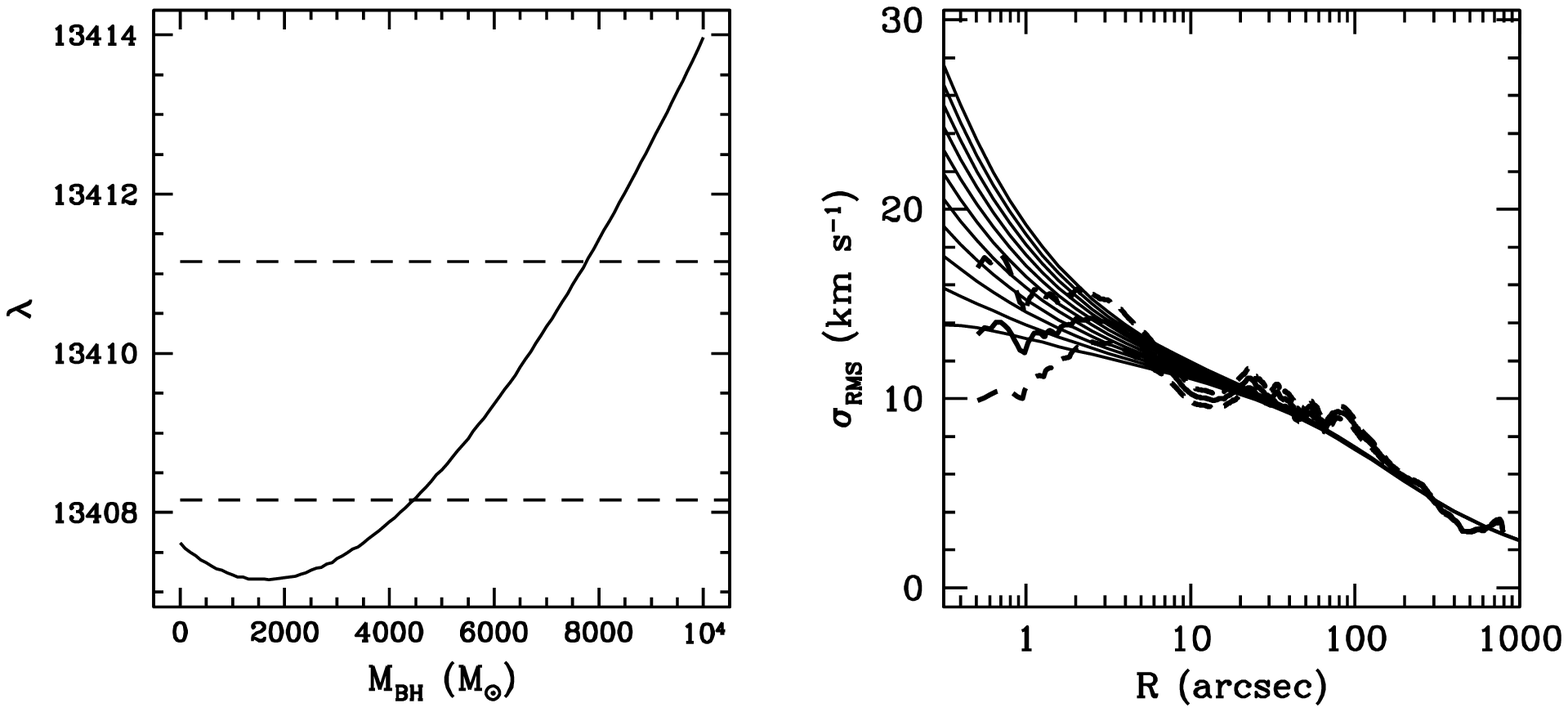}}
\ifsubmode
\vskip3.0truecm
\addtocounter{figure}{1}
\centerline{Figure~\thefigure}
\else\figcaption{\figcapmaxlike}\fi
\end{figure}


\fi 


\end{document}